%% file: main.tex
\documentclass[sigconf]{acmart}
\usepackage{multirow}
\usepackage{paralist}
\usepackage{enumitem}
\AtBeginDocument{%
  \providecommand\BibTeX{{%
    \normalfont B\kern-0.5em{\scshape i\kern-0.25em b}\kern-0.8em\TeX}}}

\copyrightyear{2022} 
\acmYear{2022} 
\setcopyright{acmcopyright}\acmConference[L@S '22]{Proceedings of the Ninth ACM Conference on Learning @ Scale}{June 1--3, 2022}{New York City, NY, USA}
\acmBooktitle{Proceedings of the Ninth ACM Conference on Learning @ Scale (L@S '22), June 1--3, 2022, New York City, NY, USA}
\acmPrice{15.00}
\acmDOI{10.1145/3491140.3528264}
\acmISBN{978-1-4503-9158-0/22/06}

\begin{document}
\fancyhead{}
\title[A Computer-aided Visualization System for Reflecting Language Learning Progress]{RLens: A Computer-aided Visualization System for Supporting Reflection on Language Learning under Distributed Tutorship}


\author{Meng Xia}
\affiliation{
  \institution{School of Computing, KAIST}
  \country{Daejeon, Republic of Korea}
}\email{mengxia@andrew.cmu.edu}

\author{Yankun Zhao}
\authornote{The authors contributed equally to this research.}
\affiliation{
  \institution{HKUST}
  \country{Hong Kong SAR}
}\email{yzhaock@connect.ust.hk}

\author{Jihyeong Hong}
\authornotemark[1]
\affiliation{
  \institution{School of Computing, KAIST}
  \country{Daejeon, Republic of Korea}
}\email{z.hyeong@kaist.ac.kr}

\author{Mehmet Hamza Erol}
\authornotemark[1]
\affiliation{
  \institution{School of Computing, KAIST}
  \country{Daejeon, Republic of Korea}
 }\email{mhamzaerol@kaist.ac.kr}

\author{Taewook Kim}
\affiliation{
  \institution{School of Computing, KAIST}
  \country{Daejeon, Republic of Korea}
 }\email{taewook@u.northwestern.edu}

\author{Juho Kim}
\affiliation{
  \institution{School of Computing, KAIST}
  \country{Daejeon, Republic of Korea}
 }\email{juhokim@kaist.ac.kr}


\renewcommand{\shortauthors}{Meng Xia et al.}

\newcommand{\eg}{{\it e.g.,\ }}
\newcommand{\etal}{{\it et al.\ }}
\newcommand{\etc}{{\it etc.}}
\newcommand{\ie}{{\it i.e.,\ }}

\newcommand{\taewook}[1]{{\color{blue} #1}}
\newcommand{\meng}[1]{{\color{red} #1}}

\begin{abstract}
With the rise of the gig economy, online language tutoring platforms are becoming increasingly popular. These platforms provide temporary and flexible jobs for native speakers as tutors and allow language learners to have one-on-one speaking practices on demand, on which learners occasionally practice the language with different tutors. With such distributed tutorship, learners can hold flexible schedules and receive diverse feedback. However, learners face challenges in consistently tracking their learning progress because different tutors provide feedback from diverse standards and perspectives, and hardly refer to learners' previous experiences with other tutors. We present RLens, a visualization system for facilitating learners' learning progress reflection by grouping different tutors' feedback, tracking how each feedback type has been addressed across learning sessions, and visualizing the learning progress. We validate our design through a between-subjects study with 40 real-world learners. Results show that learners can successfully analyze their progress and common language issues under distributed tutorship with RLens, while most learners using the baseline interface had difficulty achieving reflection tasks. We further discuss design considerations of computer-aided systems for supporting learning under distributed tutorship.
\end{abstract}





\maketitle

\input{sections/01-Introduction}
\input{sections/02-RelatedWork}
\input{sections/03-NeedsFinding}
\input{sections/04-SystemDesign}
\input{sections/05-UserStudy}
\input{sections/06-Discussion}
\input{sections/07-Conclusion}
\input{sections/00-Acknowledgement}

\bibliographystyle{ACM-Reference-Format}
\bibliography{sample-base}


\end{document}
\endinput

%% file: sections/01-Introduction.tex
\section{Introduction}
With the rise of the gig economy, temporary and flexible jobs are prevalent toward efficient resource allocation~\cite{wright2017beyond, alkhatib2018laying}. As instances of the gig economy, online language tutoring services (e.g., Cambly~\footnote{\url{https://www.cambly.com/english?lang=en}}, Preply~\footnote{\url{https://preply.com/}}, and italki~\footnote{\url{https://www.italki.com/}}) that provide part-time jobs for native speakers to work as tutors and enable language learners to have one-on-one lessons with tutors on demand are becoming increasingly popular~\cite{yeh2019speaking, kozar2014exploratory}. In contrast to the fixed instructors in the conventional language learning classroom, learners can select different tutors every learning session. The learning experience in such kind of online language tutoring services was newly identified as ``distributed tutorship'', in which learners distribute their learning time with different tutors, implying learning discontinuously in time with different tutors~\cite{xia2022understanding}. For example, in Ringle\footnote{\url{https://www.ringleplus.com/en/student/landing/home}}, a popular online English tutoring platform, 40\% of 15,959 learners change to new tutors every session; 44\% of learners change to new tutors while reverting to previous tutors sometimes; and only 16\% of learners change to new tutors and then fix on one tutor~\cite{xia2022understanding}.

\begin{figure*}
\centering
  \includegraphics[width=0.75\linewidth]{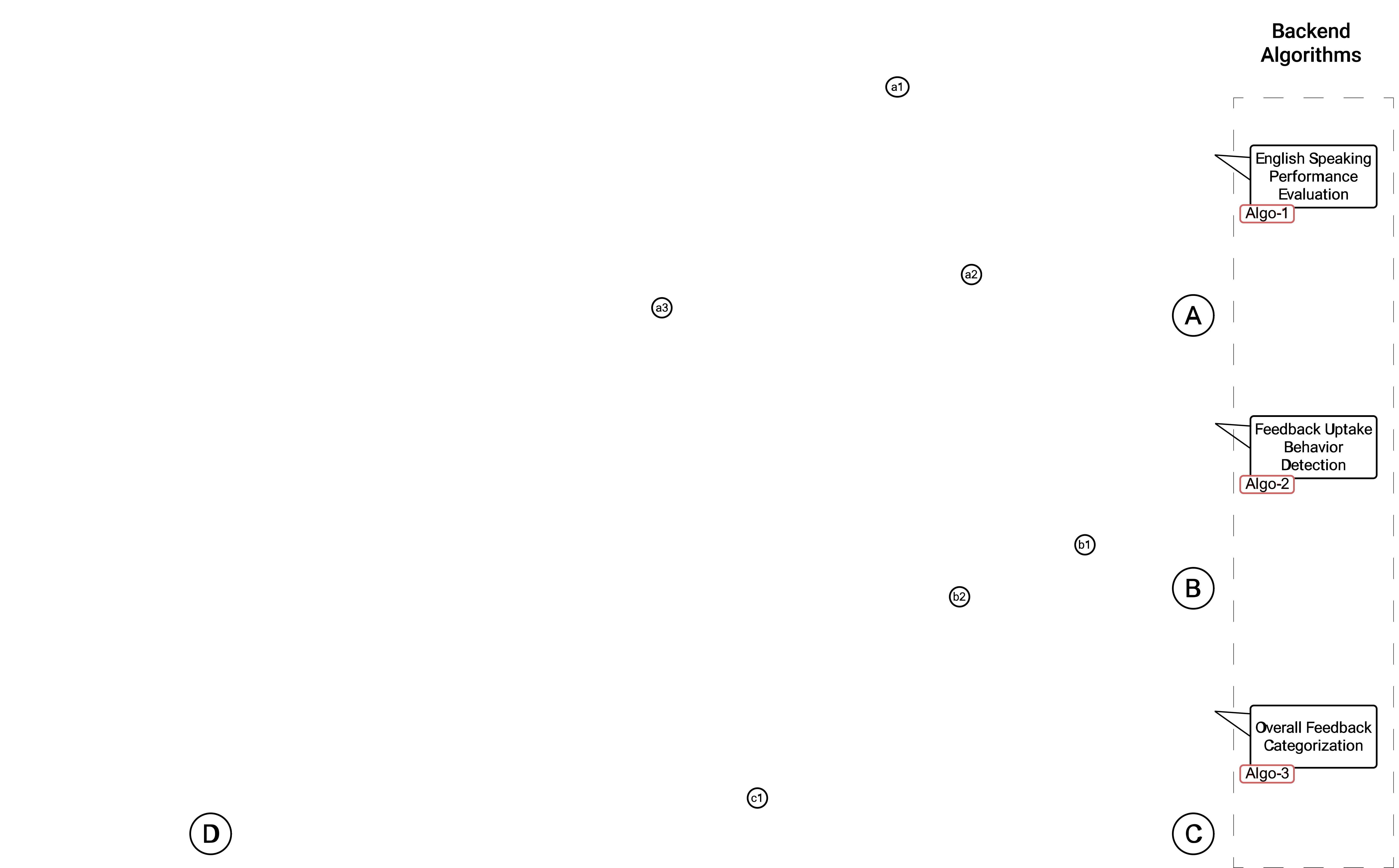}
  \caption{RLens System: Overview (A) shows the overall learning progress with both tutors' scores and system scores; Correction View (B) ranks common language issues and tracks learners' feedback uptake behaviors; Suggestion View (C) groups suggestions from different tutors; Transcript View maps tutor feedback to transcript (shown in Figure~\ref{fig:script}), and Filter Panel (D) for filtering by tutors, topics, and date. Algo1 - Algo4 are backend algorithms that drive data visualizations (Algo4 in Figure~\ref{fig:script}).} 
  \label{fig:whole_page}
  \vspace{-3mm}
\end{figure*}

Distributed tutorship brings learners convenience in scheduling tutoring sessions and benefits in receiving diverse feedback. However, higher distributedness is suggestively correlated with lower learning gains and
poses challenges for learners to reflect on their learning progress~\cite{xia2022understanding}. 
Feedback discontinuity~\cite{hattie2007power, ericsson1993role} is one of the issues in distributed tutorship.
In traditional learning, fixed instructors can provide continuous feedback to learners by pointing out their recurring bad habits or suggesting incremental improvements based on observing longitudinal learning practices. In distributed tutorship, it is hard for learners to receive such feedback since tutors have limited access and motivation to check a learner's session history with other tutors. In particular, language learning is a long process~\cite{de1978language}, where learners need to reflect on their learning practices over time to correct their common problems (e.g., tense errors, redundant filler words)~\cite{ok2014reflections, tarvin1991rethinking}. However, little research has investigated 
learners' challenges in reflecting on cumulative language learning practices under distributed tutorship and 
how computer techniques can assist the reflection process.

Previous studies proposed to improve the feedback quality by asking tutors to check the samples of other tutors' feedback before grading assignments~\cite{willey2011getting, willey2010perceived, willey2010improving}. 
However,
this process is tedious and involves privacy issues for a tutor to listen to the learner's previous audio recordings or check other tutors' feedback in online language tutoring platforms. 
To reduce tutors' workload and avoid privacy concerns, we propose a computer-aided visualization system, RLens, which utilizes natural language processing (NLP) and data visualization techniques to automatically analyze different tutors' feedback and learners' speaking transcripts for assisting learners' reflection under distributed tutorship. 

First, by interviewing 16 English learners who experienced distributed tutorship, we identified four major challenges (i.e., grading inconsistency, feedback discontinuity, unorganized feedback, lacking context for feedback understanding) that learners face in reflecting on their learning progress. We then implemented RLens to address these challenges. Specifically, to mitigate the challenge of different tutors having different grading standards, RLens calculates learners' speaking performance (Algo1) based on transcripts throughout the sessions and shows the computed scores in Overview (Figure~\ref{fig:whole_page}A).
To solve the feedback discontinuity, Correction View (Figure~\ref{fig:whole_page}B) helps learners identify common language issues by ranking the language issues pointed out by different tutors based on their frequency and recency. 
It further detects learners' feedback uptake behaviors (i.e., learners' corrective actions to the feedback~\cite{lyster1997corrective}) across sessions (Algo2) and demonstrates them using a heat map. In particular, we propose an algorithm to extract atomic corrections (e.g., suggested words) from tutors' feedback and track feedback uptake behaviors in each learning session using masked language modeling ~\cite{devlin2019bert}. Suggestion View (Figure~\ref{fig:whole_page}C) groups different tutors' suggestions using natural language inference techniques~\cite{maccartney2009natural} (Algo3) and uses a heat map to show where to focus on. Transcript View (Figure~\ref{fig:script}) helps the learner to understand the context of the feedback by mapping tutors' feedback to the transcripts based on the sentence similarity (Algo4). A Filter Panel (Figure~\ref{fig:whole_page}D) is integrated into RLens to filter tutoring sessions.

We evaluated RLens in a between-subjects study with 40 real-world learners by asking them to reflect on their actual learning data. 
Results show that learners can successfully analyze their progress and common language issues under distributed tutorship with RLens, while most learners using the baseline interface had difficulty achieving reflection tasks.
Our contributions are:
\begin{itemize}[leftmargin=*,itemsep=0pt,topsep=0pt]
    \item A computer-aided visualization system facilitating learners' reflection on the learning process under distributed tutorship.
    \item A user study showing the effectiveness of reflecting learning progress with RLens, and a set of design considerations for computer-aided learning systems under distributed tutorship.
\end{itemize}

%% file: sections/02-RelatedWork.tex
\section{Related Work}
This section reviews previous work on online language tutoring platforms, feedback quality control, and computer-assisted systems for reflection in language learning.
\subsection{Online Language Tutoring Platforms}
Gig economy has gained popularity by providing temporary and flexible jobs for efficient resource allocation and it brings new practices and opportunities in learning and teaching~\cite{xia2022understanding}. 
Online language tutoring is an emerging type of language learning in gig economy~\cite{kozar2015discursive}. This mechanism provides jobs to native speakers to work as tutors and allows language learners to have one-on-one lessons with native speakers with low time and distance barriers~\cite{kozar2014exploratory}. 
Most of previous studies have explored the basic characteristics of different stakeholders in online language tutoring platforms as opposed to the education mode. Some focused on the tutors' perspectives on the online tutoring setting~\cite{werbinska2019english, yung2020most}. 
Other studies investigated learners' demographics, goals, expectations~\cite{kozar2014exploratory}, and motivations~\cite{yung2020secondary}. For example, the analysis of 121 application forms from a private tutoring platform in Russia revealed that the majority of learners are adults, and their motivations for having online tutoring services are work-related and examination-related~\cite{kozar2014exploratory}.

Recent work investigated how the learning outcome is influenced by the new learning mode, distributed tutorship (i.e., learners occasionally practice the language with different tutors)~\cite{xia2022understanding}. It demonstrated that distributed tutorship is highly active and suggestively correlates with lower learning gains. However, most studies mentioned above are analytical in nature. We attempt to bridge the gap between these analyses and real-world learners through a system to address learning challenges under distributed tutorship.

\subsection{Feedback Quality Control with Multiple Tutors}
While reflection under distributed tutorship is not well studied, issues in feedback quality control when learning from multiple tutors have been explored before, such as grading inconsistency~\cite{willey2011getting} and feedback discontinuity~\cite{harvey2013written, xia2022understanding}. Grading inconsistency refers to inconsistent marking standards and feedback quality amongst different tutors. 
Researchers introduced SPARK, a software tool for tutors to give feedback by comparing average marks and other tutors' feedback~\cite{willey2010perceived,willey2010improving}, or an advanced version, SPARK+, to additionally support discussions among tutors to address grading inconsistency~\cite{willey2011getting}. Similar issues and methods are mentioned in the peer grading in MOOCs~\cite{luo2014peer}. 

Feedback discontinuity means that the feedback given by different tutors across learning sessions lacks coherence and does not focus on the same learning goal, as previously reported in a medical education program~\cite{harvey2013written}. The authors reported that only 16\% of the written feedback given by geographically distributed supervisors mentioned students' clinical performance over time continuously. The authors suggested giving more detailed feedback and having communication among tutors before giving feedback. Another work also mentioned that learners need to receive continuous feedback in the acquisition of expert performance~\cite{ericsson1993role}.
However, previous methods are not applicable in online language tutoring to solve the grading inconsistency and feedback discontinuity issues. 
They are tedious and not scalable~\cite{willey2011getting}, while raising privacy issues when a tutor listens to the learner's previous audio recordings or check other tutors' feedback. In addition, many tutors are part-time workers and can devote only limited time to online tutoring~\cite{werbinska2019english}. Instead of introducing more workload to tutors and avoiding privacy issues, we propose using NLP techniques to automatically organize tutors' feedback and trace learners' learning progress based on tutors' feedback.

\subsection{Computer-Assisted Systems for Reflection in Language Learning}
Computer-assisted tools have been developed to facilitate learning language skills: writing, speaking, listening, and reading. A computer-supported collaborative prewriting tool was developed for enhancing young L2 learners' writing performance~\cite{lan2015computer}. In speaking, recent work proposed using exaggerated audio-visual corrective feedback to help learners with pronunciation~\cite{bu2021pteacher}. In listening, a computer-assisted shadowing trainer was developed for self-regulated foreign language listening practice~\cite{reza2021designing}. Finally, in reading, a computer-supported ubiquitous learning environment was designed for vocabulary learning~\cite{ogata2010computer}. Some of these tools utilize NLP techniques to analyze and evaluate learners' language learning skills. They also utilize different types of visualizations to provide visual feedback to learners for reflection. However, most tools are designed for English learning at the level of a word, a sentence, or a single session instead of supporting reflection of learning progress over multiple sessions. 

Visualization could effectively present learning data to promote self-reflection~\cite{govaerts2012student, xia2020using}. The language learning process is a sequence of learning events in each tutoring session, and previous studies visualize event sequences by placing events along a horizontal time axis, such as Lifelines~\cite{plaisant1996lifelines}, CloudLines~\cite{krstajic2011cloudlines} and TimqueSlice~\cite{zhao2013interactive}. Inspired by these techniques, we propose a set of designs including a timeline-based heat map to show the learning progress. 

%% file: sections/03-NeedsFinding.tex
\section{Formative Study}
\label{sec:needs_finding}
To understand learners' practices and challenges when reflecting on their learning progress under distributed tutorship, we conducted semi-structured interviews with 16 learners on Ringle (https://www.ringleplus.com/en/student/landing/home), a popular online English tutoring platform that provides 1:1 speaking and writing sessions with native English speakers. 

\textbf{Participants}
\begin{table}[]
\caption{The session background of the 16 participants in our needfinding interviews.}
\label{tab:needsfinding_participants}
\resizebox{\linewidth}{!}{
\begin{tabular}{l|llllllllllllllll}
\hline
Participants & P1 & P2 & P3 & P4 & P5 & P6 & P7  & P8  & P9  & P10 & P11 & P12 & P13 & P14 & P15 & P16 \\ \hline
\#of sessions & 77 & 67 & 33 & 54 & 17 & 81 & 137 & 105 & 134 & 399 & 35  & 23  & 76  & 24  & 126 & 59  \\ \hline
\# of tutors & 37 & 53 & 11 & 33 & 13 & 41 & 72  & 57  & 71  & 70  & 26  & 19  & 48  & 18  & 70  & 54  \\ \hline
\end{tabular}
}
\end{table}
As shown in Table~\ref{tab:needsfinding_participants}, we selected people who have experienced distributed tutorship (i.e., more than one tutor) on the platform.
In total, 16 learners (four males, 12 females; ages between the 20s and 50s) whose first language is not English participated. 
Their educational background ranged from having college credits to having a master's degree. There were 12 participants who reported their self-evaluated speaking skills: eight were intermediate level, and four were advanced level. Participants received USD 20 each as compensation for participating in a 60-minute interview.

\textbf{Interview Questions and Analysis Procedure}
The interview is semi-structured and questions include but are not limited to: 
(1) How often do you check tutor feedback? (2) What do you review and how long does it take?  (3) How do you evaluate yourself when feedback is given by different tutors? (4) How do you calibrate your progress when you have taken multiple lessons with different tutors? Have you encountered any difficulties? (5) If we design an interface to help you review your learning history across sessions, what functions would you want to have? Two of the authors transcribed and analyzed the interview results using content analysis~\cite{Mayring2015}. 
Key findings are summarized as follows.

\textbf{Challenges of Reflection under Distributed Tutorship}
\textit{C1: Grading inconsistency.} Based on the tutor scores, learners have difficulty knowing whether their performance has improved since they think scores provided by different tutors are of different standards.
We found that although Ringle provides grading criteria, 
all 16 participants still mentioned this situation.
\textit{``...the scores depend on the tutors too much, so they can't be a clear evaluation metric. 
'' (P13)}. 

\textit{C2: Feedback discontinuity.} Learners are uncertain whether they have applied what they had learned and unaware of their common errors, given that detailed corrections provided by different tutors are not tracked and mentioned in subsequent learning sessions. 15 participants preferred having sessions with fixed tutors for receiving continuous feedback. For example,
\textit{``with the tutor I have seen, I can have a conversation that follows the previous conversation, and they know more about my common mistakes and habits, so they know more about in which aspect I have improved, so I wish to meet same tutors again.'' (P5)}

\textit{C3: Unorganized feedback.} Learners cannot easily organize the feedback to know which aspect to focus because feedback provided by different tutors spans different perspectives. Five participants mentioned how feedback includes diverse perspectives. 
\textit{``However, the feedback is written by the tutor, so it's very subjective. Some tutors select one or two things they think are important and write it in the feedback; some tutors divide the feedback into five metrics such as grammar and vocabulary.'' (P6)} 

\textit{C4: Lacking context for feedback understanding}. Since the tutor feedback is given using a separate documentation, learners cannot easily interpret detailed feedback without context information. Six participants mentioned that they need to refer to the transcript and audio to understand the tutor's feedback. However, it is hard to find the language issue mentioned by the tutor in the transcript.


\textbf{Design Requirements} We derived design requirements based on the challenges and existing literature on continuous feedback~\cite{xia2022understanding,ericsson1993role}.

\textit{R1: Provide a data-driven assessment on learning performance along with tutors' scores over time.} 
To address the grading inconsistency (C1), we propose to provide a computed score of learners' performance based on their learning data (e.g., audio-to-text transcription)  in addition to tutors' scores.

\textit{R2: Identify common language issues and track feedback uptake behavior.} 
To address feedback discontinuity (C2), since it is difficult to have a fixed tutor in the online tutoring system~\cite{xia2022understanding}, we propose to apply NLP techniques to track and rank their language issues, rank them to find common ones, and detect learners' feedback uptake behaviors. 
For example, the system can track how the learner apply the suggested vocabulary after the tutor's correction.

\textit{R3: Organize tutor feedback automatically into different categories.} To address unorganized feedback (C3), we propose to group feedback into different categories automatically to highlight the focus area and common suggestions based on their frequency. 

\textit{R4: Map the tutor feedback to transcripts.} To facilitate learners to understand the correction within its usage context (C4), for each correction pointed by the tutor, we propose highlights and edits of the correction in the transcript. 

\textit{R5: Provide intuitive visualizations to present the learning progress (C1-C4).} Visualization could effectively present learning data to promote self-reflection~\cite{govaerts2012student, xia2020using}. In addition, five participants wanted a visual representation of their learning progress. 

%% file: sections/04-SystemDesign.tex
\section{System Design}

\label{sec:visualization_system}
To meet these requirements, we designed an interactive visualization-based system powered by a set of data-driven algorithms. 

\subsection{Algorithms for Evaluating Language Learning Progress}
\label{sec:alg}
We propose four algorithms (Algo1-Algo4) to satisfy the four requirements (R1-R4) and drive the visual interface (Figure~\ref{fig:whole_page}, Figure~\ref{fig:script} based on existing research and discussions with three experienced tutors (with an average of three years of experience) from Ringle. These algorithms take tutors' written feedback and learners' speech transcripts as input sources.
For each tutoring session on Ringle, learners receive an audio recording, audio-to-text transcription, scores on English speaking performance given by the tutor, and the tutor's written feedback. The tutor's written feedback usually contains the overall feedback and in-depth corrections. 

\textbf{Algo1: English Speaking Performance Evaluation}
\label{sec:caf}
To evaluate the English speaking computationally (R1), we adopt the metrics proposed in previous English education research~\cite{housen2009complexity, skehan1998cognitive, ellis2003task, Ellis2005AnalysingLL}: complexity, accuracy, and fluency. Since there is no fixed and optimized measurement for each metric, we select commonly used measurements from the literature. In particular, we calculate \textit{Vocabulary Complexity} based on the measure of textual lexical diversity (MTLD) ~\cite{mccarthy2005assessment}, which calculates the average length of sequential words a speaker can produce that keep the type-token ratio (TTR) higher than $x$. $x$ is set to $0.72$ by referring to ~\cite{mccarthy2010mtld, fergadiotis2015psychometric}. TTR is the ratio of the number of different words (i.e., types) to the total number of words (i.e., tokens) ~\cite{colman2015dictionary}. \textit{Grammar Accuracy} is calculated as the ratio of error-free C-Units to the total number of C-Units, where C-Unit is defined as the minimal communication unit (e.g., ``Yes.'') ~\cite{foster2000measuring, hughes1997guide}. In terms of \textit{Fluency}, we calculate the Mean Length of Run ~\cite{iwashita2008assessed}, the average number of syllables per utterance without any pause, where the threshold for pause identification is set to $250 ms$ in accordance with previous cases~\cite{kahng2014exploring, prefontaine2016utterance}.

\textbf{Algo2: Feedback Uptake Behavior Detection}
Feedback uptake behavior refers to learners' corrective actions according to tutors' feedback~\cite{lyster1997corrective}. 
We focus on the feedback uptake behavior for corrective feedback (e.g., two apple -> two apples) because uptake behavior for high-level feedback (e.g., watch more English movies) is difficult to track through speech data without additional resources. Tutors' written corrective feedback on Ringle contains the original sentence spoken by the learner, the corrected sentence by the tutor, and the correction type, namely grammar, vocabulary, and fluency. 
To detect feedback uptake behavior (R2), we propose the following pipeline: (1) detecting which language issue is pointed out by the tutor (e.g., a subject-verb disagreement in grammar); (2) detecting whether learners still have this issue or corrected it in their subsequent speaking sessions. We introduce how we detect feedback uptake behaviors for each correction type as follows.

\textit{Grammar.} 
For each grammar error pointed out by the tutor, we consider grammar errors of the same type in learners' subsequent sessions in detecting the feedback uptake behavior. To this end, we first detect which type of grammar error is pointed out by the tutor by comparing the original sentence and the corrected sentence using an open-source grammar checker~\footnote{\url{https://github.com/languagetool-org/languagetool}}). 
Second, 
we count the number of that type of error in subsequent lessons using the grammar checker. In particular, we fine-tuned the grammar checker to suit the characteristics of spoken English and tolerate auto speech recognition errors by ignoring the grammar errors caused by capitalization, punctuation, homophones, repeats, pauses, false starts, corrections, interjections, and stutters~\cite{fadhila2013errors}. 

\textit{Vocabulary.}
\label{sec:feedback_uptake}
For vocabulary corrections, we consider two types of feedback uptake behaviors. The first type is forgetting to apply the suggested expression (vocabulary or phrase), in which the learner used the original expression when the suggested one can be used. The second type is applying the suggested expression correctly. For example, for a pair of original and suggested expression: ``request'' and ``require'', the first type of uptake behavior will be detected if the learner said ``The job requests at least two years of related experience.'' And the second type of uptake behavior is detected if there is a sentence like ``This document requires your signature.'' in the speaking transcripts.

To achieve this goal, we first extract the pair of original and suggested expressions given the original and suggested sentences. The detailed steps are as follows. (1) Find out the word differences between two sentences using ERRANT~\cite{bryant-etal-2017-automatic, felice-etal-2016-automatic} as the fundamental algorithm. For example, suppose that the original sentence is ``She always tries to think positively.'' and the suggested sentence is ``She is always so optimistic.'' Then the difference found is changing ``tries to think positively'' to ``is so optimistic''. (2) Enumerate two lists of possible expressions from different parts in the original and suggested sentence respectively, where each expression must contain at least one of a noun, verb, adjective, or adverb. For example, we have list 1 from the original sentence: \{``tries'', ``think'', ``positively'', ``tries to'', ``to think'', ``think positively'', ``tries to think'', ``to think positively'', ``tries to think positively''\}, and list 2 from the suggested sentence: \{``optimistic'', ``so optimistic'', ``is so optimistic''\}. (3) Pick one expression from each list to form a pair and check the expressions in which pair has the most similar meaning in context using a sentence transformer (MPNet-base-v2 model~\cite{song2020mpnet, reimers2019sentencebert}). After matching, ``think positively'' and ``optimistic'' is the pair that turns out to be the most similar, hence is extracted.

Second, we detect the feedback uptake behavior (i.e., forgetting to apply the suggested expression or applying the suggested expression correctly) in the subsequent lessons. Given that the use of vocabulary is highly dependent on the contextual semantics, we utilize the ALBERT-xxlarge-v2 model~\cite{lan2020albert, wolfetal2020transformers}  pretrained using the masked language modeling (MLM) objective~\cite{devlin2019bert}, which allows the model to fuse both of the left and right context of a masked word. The model thus can take contextual semantics into consideration.
Specifically, for a pair of original and suggested words, once we find the original word (or its variants/derivatives) in transcripts, we first mask the word, then use the model to predict the masked token in the sentence. If the suggested word (or its variants/derivatives) is one of the predicted words, we identify this occurrence of the original word as a recurring error. We apply a similar idea for detecting correct applications of a suggested word. We mask the suggested word found and see if the original word is in the predicted word candidates list. Our uptake behavior detection algorithm is evaluated on a random sample (10\%, 126 sentences) from the corpus of written feedback in 20 learners' data. This evaluation set is labeled by three experienced tutors (three years of experience on average). Our detection algorithm shows the precision of 91.49\% and the recall of 92.47\% for detecting vocabulary uptake behaviors.

\textit{Fluency.}
Since there is hardly consensus on fluency evaluation~\cite{rossiter2009perceptions}, we chose the frequency of filler words to show the fluency feedback uptake behavior, which is pointed out as one of the most common issues by the three experienced tutors. We adopt five common filler words from previous work: ``uh'', ``um'', ``like'', ``you know'', and ``I mean'' ~\cite{charlyn2014um} and detect them in learners' transcripts. An occurrence of ``uh'', ``um'', or ``like'' is counted as a filler word when its part of speech is an interjection. ``You know'' or ``I mean'' is counted as a filler word when ``know'' or ``mean'' only has a subject but no object.

\textbf{Algo3: Overall Feedback Categorization}
\label{sec:feedback_group}
We categorize each sentence of the overall feedback from different tutors to help learners easily know where to focus based on the frequency of different categories (R3). To determine the categories, we first ran topic modeling using BERTopic ~\cite{grootendorst2020bertopic} on around 30,000 feedback sentences, and we identified 49 topics as our candidate categories. Then, with the help of three tutors, we manually selected seven categories: ``grammar'', ``vocabulary'', ``pronunciation'', ``fluency'', ``sentence structure'', ``compliment'', and ``greeting'', considering how useful and actionable the feedback is as the main criteria. We further combine ``compliment'' and ``greeting'' to ``other'' category. Finally, we use an ensemble of a pretrained RoBERTa-large-MNLI model~\cite{liu2019roberta, wolfetal2020transformers}, a pretrained XLM-RoBERTa-large-XNLI model~\cite{conneau2019unsupervised, wolfetal2020transformers}, and a pretrained XLM-RoBERTa-large-XNLI-ANLI model~\cite{conneau2019unsupervised, wolfetal2020transformers} to perform sentence classification for each feedback sentence. The classification accuracy of our ensemble model on a labeled sampled (575 sentences, 10\% of all the data used in the user study in section ~\ref{user_study}) tutors' feedback corpus is 
74.09\%, where the corpus is labeled by three authors.

\textbf{Algo4: Feedback-Transcript Mapping}
To match a sentence pointed by the tutor to a sentence in the transcript (R4), we utilize the similarity detection function from an NLP library, spaCy (https://spacy.io).
\vspace{-3mm}
\subsection{Visualization System}
Leveraging the algorithms in Section~\ref{sec:alg}, we designed visualizations of RLens (Figure~\ref{fig:whole_page}) to present the learning progress (R5).

\textbf{Overview}
Overview (Figure~\ref{fig:whole_page}A) is designed to help learners understand their overall learning progress by presenting both tutors' scores and computed scores calculated by our algorithms (Sec.~\ref{sec:caf}). We show both scores because tutors' scores are meaningful to some degree but may be not consistent, and computed scores can be used as a reference in that they are generated according to same criteria. Since the two scores use different schemes and require different y-axes, we juxtapose two line charts for comparison instead of superimposing them~\cite{gleicher2011visual}. The purple line represents the tutors' scores, and the yellow line represents the computed scores. Two line charts share the x-axis to show the session number.
Learners can track the tutor change to evaluate scores associated with the classes of some tutors, or the distributedness of their tutorship by observing the change of the white and grey background of the line chart. We also use a window (Figure ~\ref{fig:whole_page}A\_a1) with regression lines to provide learners with an overall trend of their scores in all sessions. The bar chart (Figure ~\ref{fig:whole_page}A\_a3) between two sessions shows the time gap (days). 

\textbf{Correction View}
Correction View (Figure ~\ref{fig:whole_page}B) ranks the common language issues and demonstrates the feedback uptake behavior for learners to prioritize what aspects of their language learning should be improved further. As introduced in Algo2 in Section~\ref{sec:feedback_uptake}, feedback uptake behaviors are analyzed from three aspects: grammar errors, vocabulary advice, and fluency suggestions. For each aspect, we use an independent tab to display the corresponding information. For example, as shown in the \textbf{Grammar Mistakes} tab in Figure~\ref{fig:whole_page}B, grammar errors are grouped into different categories and ranked based on severity. The \textbf{severity} is calculated based on the prediction of the error count in the next session using the regression line to simulate the trend of the error count by considering recentness and frequency (i.e., the more recent and more frequent error type is ranked higher).
To help learners to perceive the overall trend of the grammars errors across sessions, we visualize the frequency of errors using a heat map - a widely used, effective, and simple visual technique in showing the frequency of learning activities~\cite{37jugo2015integrating, 54xia2018instructor, 65mazza2007coursevis, 62mazza2004visualising}.
Tiles with a higher frequency are encoded with a darker shade of red.
We use a short line ``-'' in the tile to indicate that no tutor has pointed out the error yet. 
In addition, if a learner wants to check the issues pointed out by the tutor without the feedback uptake detected by the system, they can disable the feedback uptake toggle beside the drop-down menu. 
\begin{figure}
\centering
  \includegraphics[width=0.9\columnwidth]{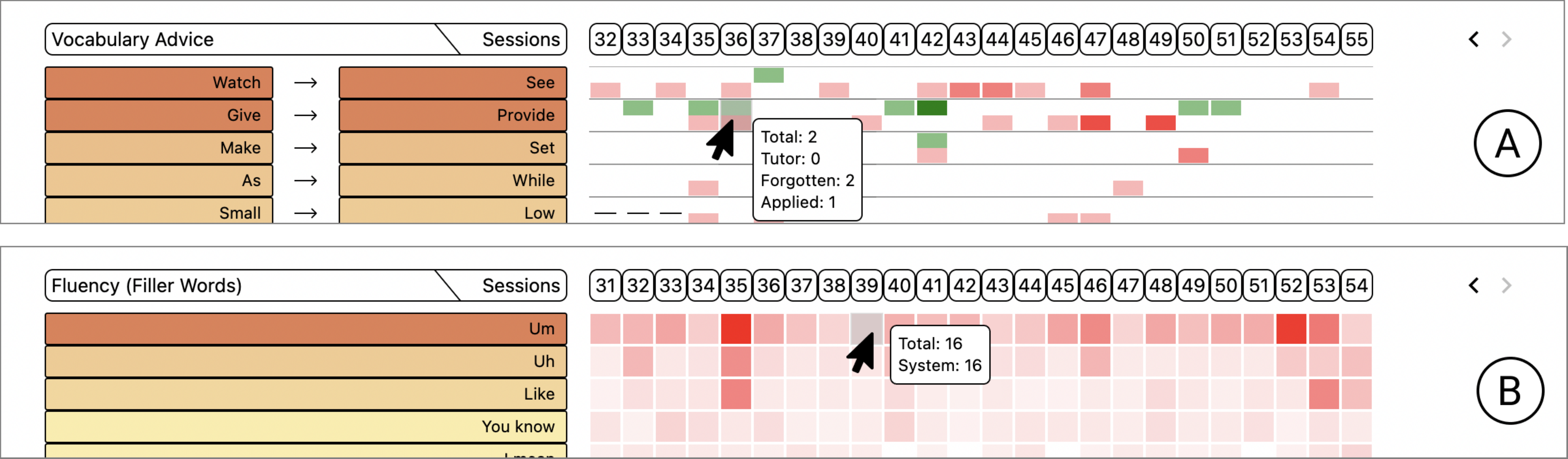}
  \caption{Corrections View: Vocabulary Advice Tab (A) and Fluency (Filler Words) Tab (B)} 
  \label{fig:vocabulary_and_fluency}
\end{figure}

The \textbf{Vocabulary Advice} tab, shown in Figure~\ref{fig:vocabulary_and_fluency}A, lists the vocabulary advice based on their severity (estimation of the times the user forgot to apply the suggested expressions in the next session). The format of vocabulary advice is ``original word - suggested word''. Different from the grammar errors, we use a two-color heat map for each vocabulary advice to show how learners address that feedback in one learning session. The shade of green of the top-half of a tile indicates how frequently that learner applied the suggested expression correctly; the shade of red of the bottom-half of a cell indicates that how frequently that learner used the original expression in the learning context when the suggested expression can be applied instead of the original one.
For example, in Figure~\ref{fig:vocabulary_and_fluency}A, the learner applied vocabulary advice "provide" correctly once and used the original words without applying the suggested word twice in session 36. 
In the \textbf{Fluency (Filler words)} tab shown in Figure~\ref{fig:vocabulary_and_fluency}B, the system lists the common filler words (i.e., uh, um, I mean, like, you know) and ranks them based on their frequencies. 

\begin{figure}
\centering
  \includegraphics[width=0.9\columnwidth]{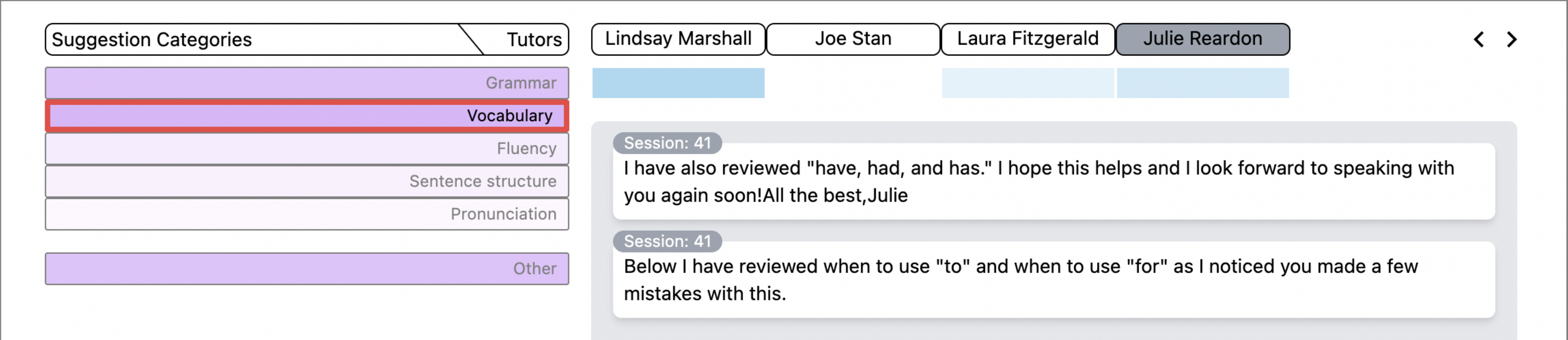}
  \caption{Suggestions View (when clicked)} 
  \label{fig:suggestion}
\end{figure}

\textbf{Suggestion View}
Suggestion View (Figure ~\ref{fig:whole_page}C) groups all the tutors' suggestions into six categories~\ref{sec:feedback_group}.
To help learners prioritize the suggestions, the system ranks all the categories based on the number of sentences mentioned by tutors in each category. Moreover, we use a heat map for each category to show each tutor's contribution (i.e., number of sentences) to that category. The darker the blue, the more sentences are mentioned by a tutor for that category. 
When the learner clicks a tile, the sentences belonging to the category and tutor are shown in a pop-up (Figure~\ref{fig:suggestion}).
\begin{figure}
\centering
  \includegraphics[width=0.9\columnwidth]{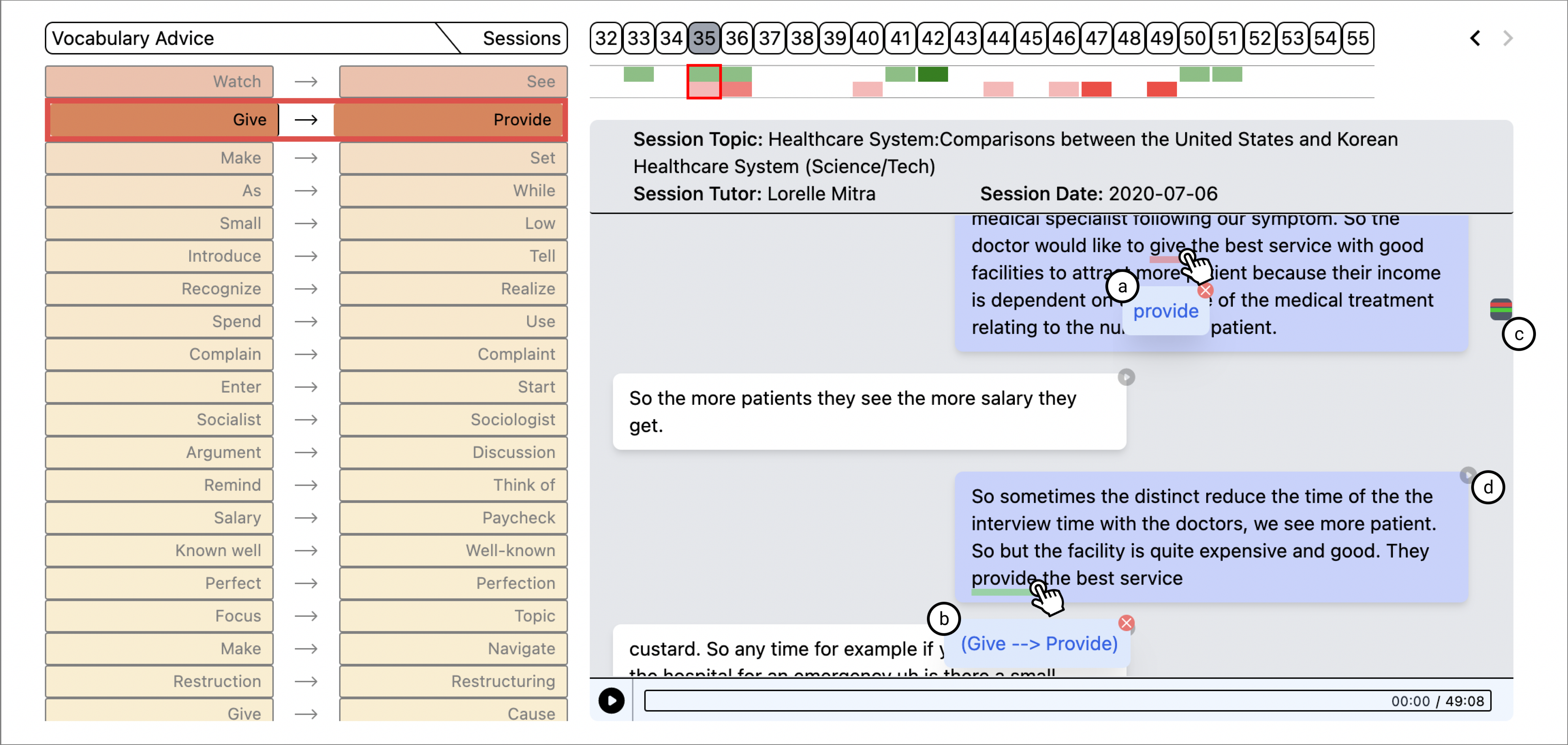}
  \caption{Transcript View with Algo4 the driven algorithm} 
  \label{fig:script}
  \vspace{-5mm}
\end{figure}

\textbf{Transcript View}
Transcript View is designed to understand the learning context by mapping the correction to the transcripts. For example, as shown in Figure~\ref{fig:script}, when the learner clicks vocabulary advice in Correction View (e.g., Give -> Provide) and further clicks session 35, Transcript View will pop up to show the feedback uptake locations in the session's transcript. In particular, for an immediate localization of the errors, it uses a red and green bar (Figure~\ref{fig:script}\_c) over the scrollbar to show where the feedback uptake is in the transcript. It also highlights places in the transcript where a learner forgot to apply the suggested vocabulary using red underlines (Figure~\ref{fig:script}\_a) and applied the suggested vocabulary correctly using green underlines (Figure~\ref{fig:script}\_b). Learners can check the corrected expression by clicking the underlined error(Figure~\ref{fig:script}\_a). 
Meanwhile, learners can click the play button (Figure~\ref{fig:script}\_d) to listen to the audio of each turn in the conversation to recall the learning context.

\textbf{Filter Panel}
RLens provides a Filter panel (Figure ~\ref{fig:whole_page}D) from different perspectives (e.g., tutors, topics, sessions, dates, etc.).

%% file: sections/05-UserStudy.tex
\section{Evaluation}
\label{user_study}
We evaluated the usefulness and efficacy of RLens in assisting learning progress reflection under distributed tutorship. 

\subsection{Study Design} 
\begin{figure}
\centering
  \includegraphics[width=0.75\columnwidth]{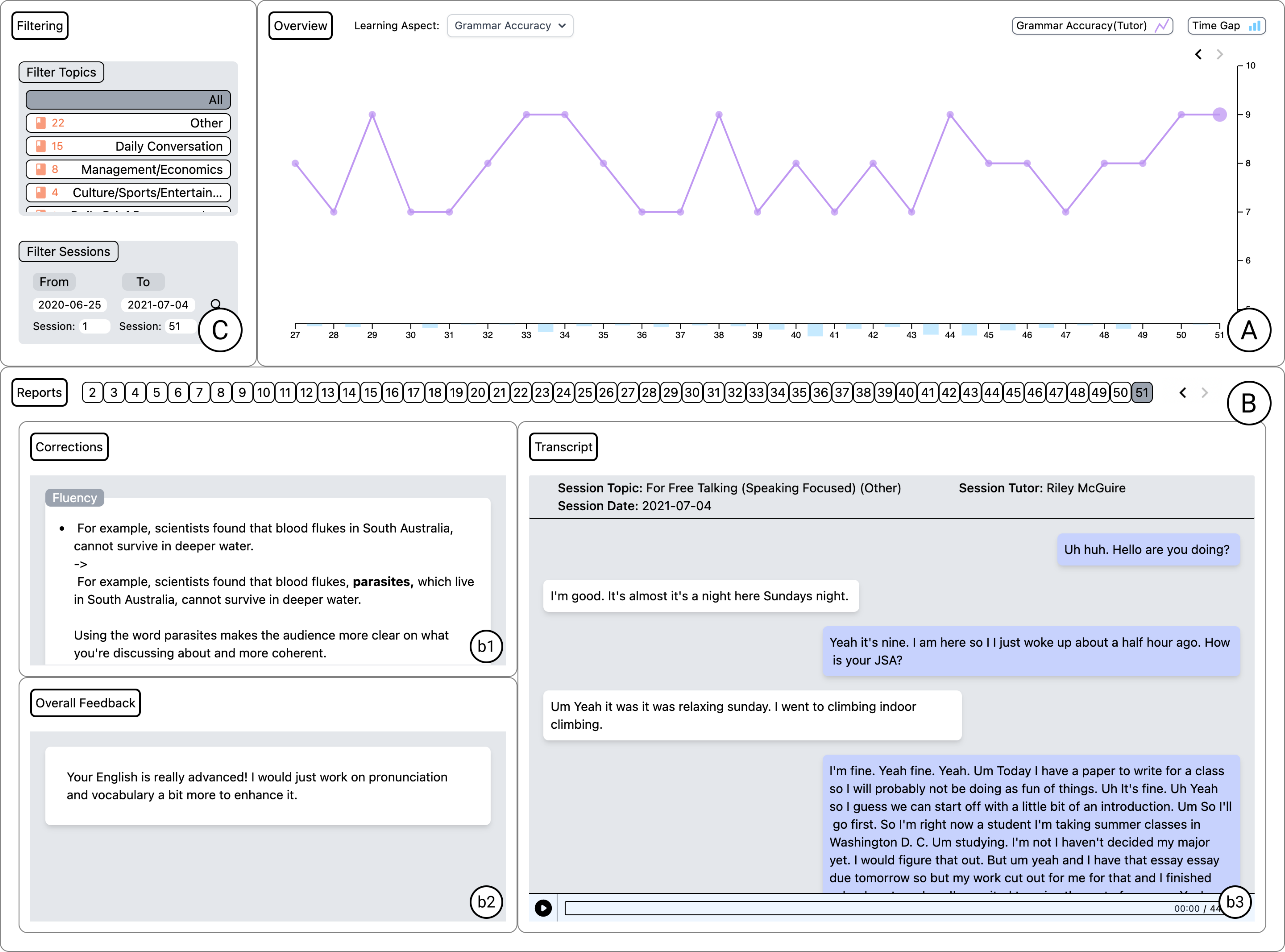}
  \caption{Baseline System:Overview (A), Report View (B), and Filter Panel (C)} 
  \label{fig:baseline}
  \vspace{-3mm}
\end{figure}
We conducted a between-subjects study on a Baseline system and RLens. Since there is no support for distributed tutorship in previous systems, we constructed a Baseline system (Figure~\ref{fig:baseline}) to simulate the learners' dashboard currently provided by the Ringle platform and other language tutoring platforms
~\cite{kozar2015discursive, xia2022understanding}: the tutor scores (Figure~\ref{fig:baseline}A), the tutor feedback (Figure~\ref{fig:baseline}B\_b1, b2), and the speaking transcript (Figure~\ref{fig:baseline}B\_b3).
We did not use Ringle directly because it contains additional information (e.g., tutors' pictures, advertisements) other than the experimental variables that might confound our
results.
We applied the same UI elements (e.g., layout, fonts) to both systems and tried to minimize visual and usability differences between them. We acknowledge that Baseline is an overly simple system, and therefore our goal is not to see if RLens beats the Baseline but rather to understand and analyze how people use RLens in depth, in comparison with the Baseline.

\textbf{Tasks} We ask people to look at their progress data and try to make sense of it for reflection. In particular, we derived seven reflection tasks from our needfinding stage and previous research on language learning reflection~\cite{yamashita2012wheel}. \textbf{T1:} Please describe your overall learning progress. \textbf{T2:} Please identify your common language issues in the English learning process. \textbf{T3:} Please describe whether you have corrected your common language issues in the learning process. \textbf{T4:} Please describe the common aspects in tutors' overall feedback.
\textbf{T5:} Please describe how you check the transcript using the system for learning. \textbf{T6:} Please describe the reasons for ups and downs in scores showing in Overview.
\textbf{T7:} Please describe how you will use this system in learning reflection if it is deployed. To simulate real reflection scenarios, we required participants to reflect on their own learning data by loading their session data from Ringle to the system upon participants' consent. 

\textbf{Measures}
We evaluate the effectiveness, informativeness, usability, and intuitiveness of RLens, refering to Weibelzahl's work~\cite{weibelzahl2001evaluation}, where the authors proposed the evaluation pipeline of interactive systems. Moreover, since people's trust and perceived accuracy is an important metric in an AI-infused system~\cite{dietvorst2015algorithm}, we also evaluate learners' trust in information provided in RLens. The questionnaire can be seen in Table~\ref{tab:questions}.

\textbf{Participants} We recruited learners from Ringle by posting an advertisement on the platform's website. 
We eliminated learners with fewer than 25 sessions to guarantee sufficient experience on distributed tutorship. Furthermore, we paired learners based on the number of sessions they had and their session/tutor ratio (i.e., \# of sessions/ \# of tutors) to guarantee that participants in both Baseline and RLens groups have a similar learning experience and distributed tutorship. Finally, we had 40 (12 males, 28 females) participants, with 20 in each group (B1-B20 in Baseline and A1-A20 in RLens). All participants' first language was not English. 
Baseline group (7 male, 13 female) had a mean age of 33.5 (min 25, max 53), a mean number of sessions of 67.05 (min 31, max 145), a mean session-tutor ratio of 1.78 (min 1.05, max 3.22), and the distribution self-reported English speaking proficiency is low (3), good (6) and intermediate (11). RLens group (5 male, 15 female) had a mean age of 35.25 (min 27, max 52), a mean number of sessions of 65.85 (min 27, max 185), a mean session-tutor ratio of 1.91 (min 1.23, max 4.56), and the distribution self-reported English speaking proficiency is low (2), good (10) and intermediate (8).
The recruitment and user study procedures were approved by the Institutional Review Board (IRB) at our university, and each participant received approximately USD 38 as compensation for participating in a 90-minute study session.


\textbf{Procedures} The user study was conducted remotely through Zoom. It contained five steps and lasted around 90 minutes: First, we introduced the background of the study, and participants read and signed the consent form.
We then introduced the interface. Participants were asked to explore the system and complete seven learning reflection tasks using the think-aloud strategy for about 40 minutes. Upon task completion, participants completed a questionnaire with 7-point Likert questions derived from existing literature~\cite{xia2019peerlens}. Lastly, we asked debriefing questions about their opinions on the most or least helpful features and suggestions for the system.

\textbf{Hypotheses} Based on previous evaluation~\cite{verbert2020learning, xia2019peerlens} of learning dashboards, we present the following hypotheses:

\textbf{H1}: RLens is more effective in helping learning progress reflection under distributed tutorship than Baseline. Specifically, RLens is more helpful for learners to clearly understand the learning progress (\textit{H1a}), be aware of common errors (\textit{H1b}) and correction behaviors (\textit{H1c}), organize tutors' suggestions (\textit{H1d}), understand learning context (\textit{H1e}), and analyze learning progress (\textit{H1f}). Therefore, learners are more willing to recommend RLens to others (\textit{H1g}).

\textbf{H2:} The information for learning progress reflection in RLens is more accessible (\textit{H2a}) and sufficient (\textit{H2b}) than Baseline for learning progress reflection under distributed tutorship.




\begin{table}
\small
\centering
\caption{A questionnaire was designed to cover five aspects: effectiveness (Q1-Q7), informativeness (Q8-Q9), usability (Q10-Q11), visualization \& interaction (Q12-17), and trust (Q18-Q21). All are 7-point Likert scale questions. Q14-Q21 are only applicable to RLens.}
\label{tab:questions}
\begin{tabular}{l} 
\toprule
Questions                                                                                                                                                             \\ 
\hline
\begin{tabular}[c]{@{}l@{}}Q1: The scores in the system help me to have a clear understanding \\of whether I have improved under distributed tutorship.\end{tabular}  \\
\begin{tabular}[c]{@{}l@{}}Q2: The system helps me in being aware of my common \\language issues under distributed tutorship.\end{tabular}                            \\
\begin{tabular}[c]{@{}l@{}}Q3: The system helps me to know whether I have corrected my \\common errors under distributed tutorship.\end{tabular}                      \\
\begin{tabular}[c]{@{}l@{}}Q4: The system helps me to organize different tutors' suggestions \\for future guidance.\end{tabular}                                      \\
\begin{tabular}[c]{@{}l@{}}Q5: The system helps me to know the learning context of tutors' \\feedback under distributed tutorship.\end{tabular}                       \\
\begin{tabular}[c]{@{}l@{}}Q6: The system helps me to analyze and reflect on my learning\\~progress under distributed tutorship.\end{tabular}                         \\
\begin{tabular}[c]{@{}l@{}}Q7: I would like to recommend this system to others \\if they learn from different tutors.\end{tabular}                                    \\ 
\hline
\begin{tabular}[c]{@{}l@{}}Q8: The information needed is easy to access to reflect on\\my learning progress\end{tabular}                                              \\
\begin{tabular}[c]{@{}l@{}}Q9: The information is sufficient to reflect my learning progress\\~under distributed tutorship.\end{tabular}                              \\ 
\hline
Q10: It is easy to learn the system.                                                                                                                                  \\
Q11: It is easy to use the system.                                                                                                                                    \\ 
\hline
Q12: Overall, the visualization designs in the system are intuitive.                                                                                                  \\
Q13: Overall, the interactions in the system are intuitive.                                                                                                           \\
Q14: The visualization in Overview is intuitive.                                                                                                                      \\
Q15: The visualization in the Correction View is intuitive.                                                                                                           \\
Q16: The visualization in the Suggestion View is intuitive.                                                                                                           \\
Q17: The visualization in Script View is intuitive.                                                                                                                   \\ 
\hline
Q18: I trust the computed score provided by the system.                                                                                                               \\
Q19: I trust the feedback uptake behavior detected by the system.                                                                                                     \\
Q20: I trust the grouping of suggestions by the system.                                                                                                               \\
Q21: I trust the mappings for the corrections to the script.                                                                                                          \\
\bottomrule
\end{tabular}
\end{table}

\subsection{Results and Analysis}

Two authors analyzed users' interactions in the tasks, verbal reasons for the ratings, and post-study interviews. We performed the Mann-Whitney U (rank) test on the questionnaire items (Q1-Q9) to test statistically significant differences between the conditions.

\textbf{H1. Effectiveness} Overall, the participants thought RLens is significantly more effective in helping with learning progress reflection under distributed tutorship than Baseline. 

\textit{Clear understanding of learning progress.} Participants reported that scores (including system scores and tutor scores) presented in RLens ($Mean = 5.95, SD = 1.504$) are more helpful for them to clearly understand their learning progress than only tutors' scores in Baseline ($Mean = 4.4, SD = 1.536$). Significance has been found in the Mann-Whitney U test ($U = 87.0, p < 0.001$, \textbf{H1a supported}). Based on participants' verbal feedback in T1 (i.e., describe the overall learning progress), 15 participants in Baseline and 16 participants in RLens thought having only the tutor scores as a source does not feel like to be a representative for their actual learning progress.
14 learners in RLens group thought that computed scores provided a more objective evaluation since they fluctuated less during the tutor change. 
In addition, some participants used the two scores in a complementary manner to understand their learning progress. 
\textit{"I will trust tutor's score for fluency because I think it should be graded by a person, not a system. For grammar, I will trust the system score because it is objectively being right or wrong. (A18)"}
\begin{figure*}
\centering
  \includegraphics[width=0.73\linewidth]{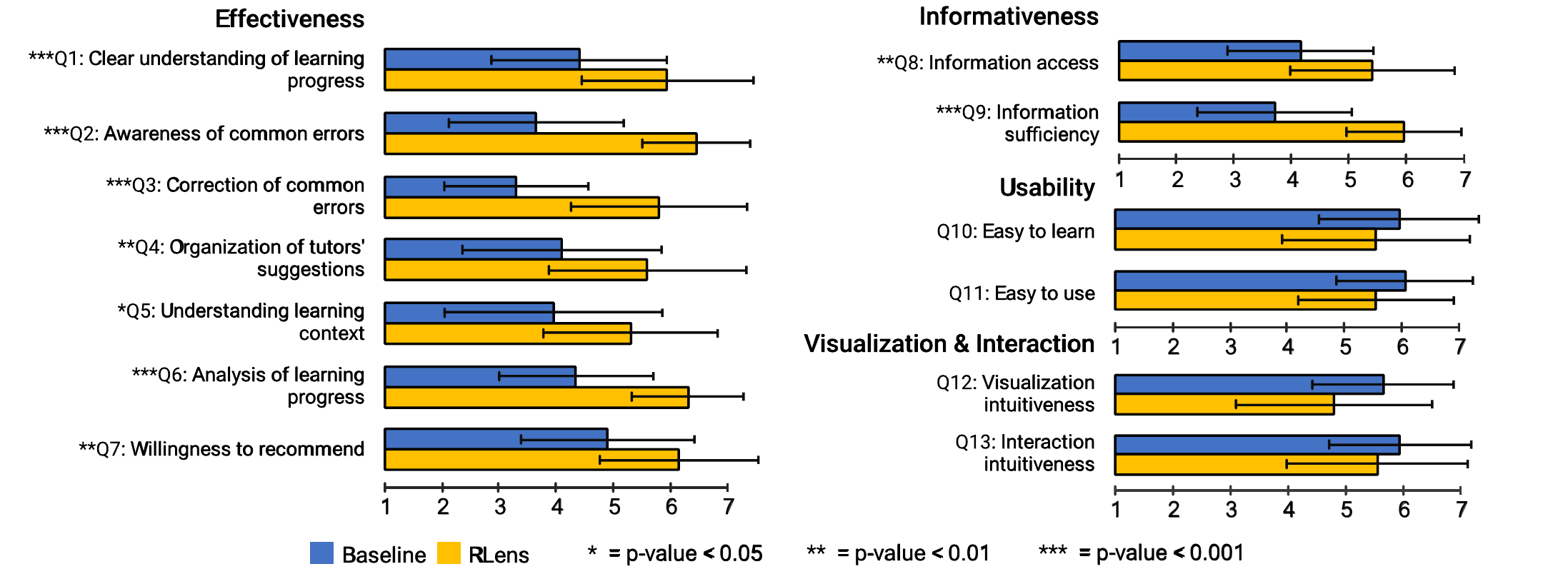}
  \caption{Means and standard errors of Baseline and RLens on effectiveness, informativeness, usability, and visualization \& interactions on a 7-point Likert scale ($*:p<.05, **: p<.01, ***:p<.001$).} 
  \label{fig:results}
\end{figure*}

\textit{Awareness of common language issues.} Participants reported they were significantly more aware of common language issues using RLens ($Mean = 6.45, SD = 0.945$) than Baseline ($Mean = 3.65, SD = 1.531$). The Mann-Whitney U test further reveals the significance ($U = 25.0, p < 0.001$, \textbf{H1b supported}). Notably, participants' answer to T2 (i.e., identifying common language issues) was extremely different between Baseline and RLens. Participants in Baseline could hardly give the answer 
by reporting that they are not sure about their common errors or 
answered the questions based on their memory in an uncertain manner. 
Many of them did not check the Baseline system since they reported that it is time-consuming to
find common errors by clicking each session. 
However, in RLens, all participants answered the question by checking different tabs in the correction view first and then confirmed their answers with their memory. Their answers pointed the specific errors, e.g., tense errors in grammar. Some participants also realized the common errors that they overlooked before, e.g., \textit{"I thought I use 'like' or 'I mean' the most, but there are lots of 'uh' and 'you know,' which I did not recognize before. (A14)"} 

\textit{Correction of common language issues.} RLens ($Mean = 5.8, SD = 1.542$) was more helpful for participants to know whether they have corrected their common errors than Baseline ($Mean = 3.3, SD = 1.261$), with Mann-Whitney U test ($U = 39.5, p < 0.001$, \textbf{H1c supported}). According to T3 (i.e., describe whether common language issues have been corrected), in Baseline, 18 participants failed to answer how they have proceeded with their common errors, and they were not sure whether they had applied tutors' corrections in their learning progress. 
Only two participants checked Correction View one by one and answered the question based on whether a specific error was mentioned by the tutor again. However, in RLens, all the participants used the heat map to answer their progress and whether they corrected their common errors. For example, A1 noted that \textit{"Filler words become lighter in the recent session, and I can see improvements in fluency. "}. RLens received positive feedback regarding the heat map that shows progress, as A8 said, \textit{"For me, it is really useful and efficient. (the color)."}. 

\textit{Organization of tutors' suggestions.} Participants reported RLens ($Mean = 5.6, SD = 1.729$) is more helpful to organize tutors' feedback for future guidance on English learning than Baseline ($Mean = 4.1, SD = 1.744$), with Mann-Whitney U test showing significance ($U = 102.5, p < 0.01$, \textbf{H1d supported}). For T4 (describe common aspects in tutors' feedback), 15 participants in Baseline had difficulties answering the common suggestion given by different tutors. 
In RLens, 19 participants answered the question with Suggestion View. One participant still felt the workload is heavy to check all sentences from one category: \textit{"It is nice to have overall feedback gathered all together, but it is hard to read it one by one. (A11)"}. 
In addition, some participants mentioned there might be conflicts in tutors' suggestions, which they wished to spot in RLens. 

\textit{Understanding learning context.} The mapping of tutor feedback to the transcript in RLens ($Mean = 5.3, SD = 1.525$) helps learners better understand their learning context than Baseline ($Mean = 3.95, SD = 1.905$). The Mann-Whitney U test shows the significance ($U = 117.5, p = 0.012 < 0.05$, \textbf{H1e supported}). Participants in Baseline found T5 (i.e., how to use the transcript) is hard. B6 said that \textit{``If I am looking at an article issue, I need to find it in the script. Then there is the cognitive effort required. ''}  
In RLens, participants all tried the mapping function by clicking the red cells and checking feedback context in the transcript view. Two participants (A12, A20) mentioned that Transcript View could be further simplified to show only sentences with errors.

\textit{Analysis of learning progress.} Participants found RLens ($Mean = 6.3, SD = 0.979$) significantly better in supporting their analysis of the learning progress under distributed tutorship than Baseline ($Mean = 4.35, SD = 1.348$). The Mann-Whitney U test confirmed the significance ($U = 43.5, p < 0.001$, \textbf{H1f supported}). In T6 (i.e., describe the reasons for learning improvements and decreases), three learners in Baseline gave up analyzing their learning progress and reported they did not have enough information. For other learners in Baseline, the major pattern for analysis was that they 
referred to their memory to explain why they had higher scores or lower scores in some classes. Participants in RLens exhibited a variety of strategies to analyze their learning progress. For example, four participants used Correction View to analyze their learning progress. Seven participants used both Overview and Correction view 
to reason about their progress. They checked Overview for the overall trend and then referred to Correction View to find the common errors. Also, A18 utilized all views to achieve the analysis. This participant first checked the lowest scores in Overview and searched the Correction View for all the errors in the corresponding session, and checked the suggestions.

Overall, participants were more willing to recommend RLens ($Mean = 6.15, SD = 1.387$) to other learners for learning under distributed tutorship than Baseline ($Mean = 4.9, SD = 1.518$). Significance is found in the Mann-Whitney U test ($U = 94.5, p = 0.002 < 0.01$, \textbf{H1g supported}).

\textbf{H2. Informativeness} Compared with Baseline, RLens received significantly higher ratings in informativeness, including information access and sufficiency. \textit{Information accessibility.} Participants found it was easier to access the information needed for learning progress reflection in RLens ($Mean = 5.4, SD = 1.429$) than Baseline ($Mean = 4.15, SD = 1.268$).
Significance has been found in the Mann-Whitney U test ($U = 90.0, p = 0.001 < 0.01$, \textbf{H2a supported}). 
\textit{Information sufficiency.} The information provided by RLens ($Mean = 5.95, SD = 0.999$) in learning reflection under distributed tutorship was perceived as more sufficient than Baseline ($Mean = 3.7, SD = 1.342$). We also observe a significant difference with the Mann-Whitney U test ($U = 43.5, p < 0.001$, \textbf{H2b supported}).
According to participants' 
reasons provided in Q8 and Q9, we found that whether having the access to tutors' information to be one of the primary reasons for the rating difference. 
\textbf{Usability}
Overall, Participants thought RLens was easy to learn ($Mean = 5.55, SD = 1.638$) and easy to use ($Mean = 5.55, SD = 1.356$). Baseline received a relatively higher score than RLens in ease of learning ($Mean = 5.95, SD = 1.395$) and ease of use ($Mean = 6.05, SD = 1.191$) while no significance was found. 

\textbf{Visualization \& Interaction}
\begin{figure*}
\centering
  \includegraphics[width=0.75\linewidth]{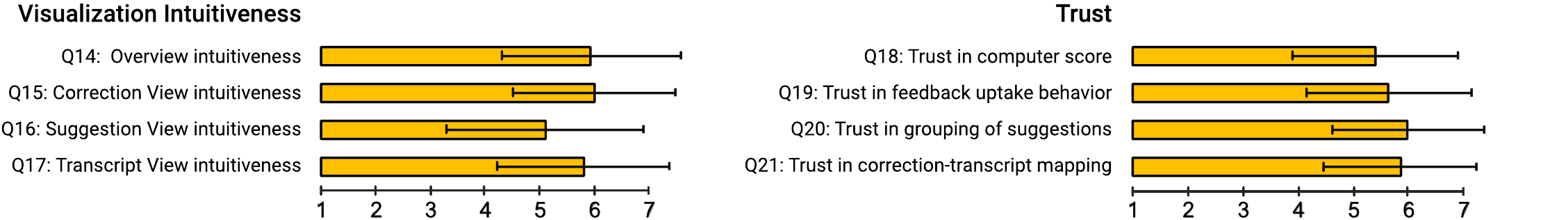}
  \caption{Means and standard errors of RLens on visualization and trust on a 7-point Likert scale} 
  \label{fig:rlens_results}
\end{figure*}
RLens received $Mean = 4.8, SD = 1.704$ for overall visualization intuitiveness, and Baseline received $Mean = 5.65, SD = 1.226$. 
Further analysis on the intuitiveness scores for each view, as shown in Figure~\ref{fig:rlens_results} (Overview: $Mean = 5.95, SD = 1.638$; Correction View: $Mean = 6, SD = 1.487$; Suggestion View: $Mean = 5.1, SD = 1.804$; Transcript View: $Mean = 5.8, SD = 1.576$) shows that the scores for individual views are relatively high, and participants in RLens reported that each view is easy to understand. However, when individual views are combined as the whole system, the large amount of information presented might have reduced the overall intuitiveness. A12 said that the whole system is complex for his age (53). 
Participants thought that the interactions in both RLens ($Mean = 5.55, SD = 1.572$) and Baseline  ($Mean = 5.95, SD = 1.234$) were intuitive. 

\textbf{Trust:} Overall, participants reported that they trust the information calculated by the algorithms. As shown in Figure~\ref{fig:rlens_results}, their ratings for trust in computed score in the Overview is $Mean = 5.4, SD = 1.501$; feedback uptake behavior in Correction View is $Mean = 5.65, SD = 1.496$; groups of tutors' feedback in Suggestion View is $Mean = 6, SD = 1.376$; and mappings for the correction to the transcript is $Mean = 5.85, SD = 1.387$.
Most participants held a positive attitude towards the computer-generated information because they did not spot errors during the exploration process. 
Some spotted system errors in the user study (e.g., "focus -> topic" was extracted from ``His answers got out of the focus.'' -> ``His answers was off the topic.''), and they rated the view where they spotted the error with a low score. However, they said this feeling did not affect their trust in other views/information in the post interviews.

%% file: sections/06-Discussion.tex
\section{Discussion}
This section discusses design considerations we learned from the study and limitations that we can address in future work. 
\subsection{Design Considerations}
\textbf{DC1: Organize information from the dimension of tutor.}
Our user study showed that learners want to access learning data with particular tutors when learning with multiple tutors. 
Some participants compared the grading standard of different tutors to understand their actual learning performance because they were worried some tutors are excessively generous or harsh. 
In addition, many participants also evaluated their learning progress with particular tutors to decide which tutor they would like to select for future learning sessions. 
A computer-aided learning system for distributed tutorship should provide information organized from the dimension of the tutors. When learning under distributed tutorship, the diversity of tutors becomes an important factor for learners to evaluate their learning progress and make future decisions on learning.
 
\vspace{2pt}\noindent\textbf{DC2: Utilize computer as a reference tutor.}
Distributed tutorship can provide flexible learning schedules and diversified feedback and language styles. 
However, tutors in distributed tutorship face challenges in maintaining the same grading standards, detecting detailed language issues, and giving continuous feedback by tracking learners' progress.
In the user study, some participants only trusted computed scores, while most participants checked and compared both human tutors' scores and computed scores. Moreover, most participants used RLens as a reference tutor to find their common language issues.
Our proposed method received favorable responses by combining high-quality feedback from different human tutors and the consistent tracking ability of the system to provide a continuous learning experience. 
We suggest that future learning tools designed for distributed tutorship utilize the computer as a reference tutor to provide data-driven assessment and give continuous feedback, complementing the role of human tutors.

\vspace{2pt}\noindent\textbf{DC3: Provide access to all information but surface actionable information.}
Participants in the user study exhibited various strategies to utilize different views to analyze their learning progress, and they appreciated the access to all the learning information from different levels. They particularly liked the design of highlighting severe language issues in the Correction View.
As pointed out by previous research, learning dashboards should provide learners with actionable suggestions~\cite{verbert2020learning, jovanovic2020supporting}. In particular, when learning with multiple tutors, different tutors point out diverse issues and give various suggestions, and it is challenging for learners to figure out the priority and distill actionable insights.
A few participants mentioned that the excessive amount of information presented in the system made it difficult to prioritize, and the current grouping algorithm of tutors' feedback in Suggestion View could be further refined to solve the conflicts of different tutors' feedback and make the suggestions more actionable.
Future learning systems for distributed tutorship should surface actionable information by distilling common feedback, resolving conflicts through algorithms, and highlighting them in the interface.

\subsection{Generalization}
Our pipeline and designs can be generalized to other platforms that might have distributed tutorship dynamics and feedback culture—for example, online skill practice (e.g., writing) using P2P skill-sharing communities (e.g., Clascity~\footnote{\url{https://clascity.com/}}), freelance markets, (e.g., Upwork~\footnote{\url{https://www.upwork.com/}}). These communities or platforms are developed to help individuals freely share their skills and receive feedback from one another. Hence, users on these platforms also potentially experience distributed tutorship. The algorithms and visualization designs of RLens can be generalized to assist users in keeping track of learning progress and organizing diverse feedback in broader domains.

\subsection{Limitations}
Our work has several limitations that have to be considered. First, the accuracy of the feedback uptake behavior algorithms and the feedback categorization algorithm can be further improved, and they are only tested on a small test set we labeled due to the lack of a labeled dataset.
Second, the multi-view dashboard of the system might impose a steep learning curve and information overload. However, as being the first system in the space to address distributed tutorship, the current interface is not meant to be a complete solution on its own but rather a prototype built to investigate how reflection on learning progress can be supported in various data-driven ways.
Different parts of our work could be simplified and adapted for different platforms and learning contexts since each view or algorithm can be used in a modular manner. 
Third, we have not introduced pronunciation metrics in RLens because the target user group has a relatively advanced speaking level and fewer pronunciation issues, while this metric is also important to be added. In addition, more language elements like grammar complexity and dialogue dynamics can be further considered.

%% file: sections/07-Conclusion.tex
\section{Conclusion and Future Work}
In this work, 
we proposed RLens, a computer-aided visualization system that allows learners to analyze and reflect their language learning progress under distributed tutorship. It utilizes NLP and information visualization techniques to empower learners to understand their learning progress and recognize common errors. We conducted a between-subjects study with 40 real-world learners. Results show that learners can successfully analyze their progress and common language issues under distributed tutorship with RLens. 
We further discuss design considerations of computer-aided learning systems with multiple tutors.
In the future, we plan to deploy RLens in real-world online language tutoring platforms to test its usability and algorithmic accuracy in the long term.

%% file: sections/00-Acknowledgement.tex
\section{Acknowledgement}
This work was supported by Institute of Information \& communications Technology Planning \& Evaluation (IITP) grant funded by the Korea government (MSIT) (No.2020-0-02237, Personalized Progress Analysis and Exercise Recommendation for Remote Language Learning Using AI and Big Data).